\documentclass[fleqn,usenatbib]{mnras}

\usepackage{newtxtext,newtxmath}
\usepackage[T1]{fontenc}

\DeclareRobustCommand{\VAN}[3]{#2}
\let\VANthebibliography\thebibliography
\def\thebibliography{\DeclareRobustCommand{\VAN}[3]{##3}\VANthebibliography}

\usepackage{graphicx}	% Including figure files
\usepackage{amsmath}	% Advanced maths commands
\usepackage{multirow}

\title[Radio light curve of PG~1302-102]{The multi-frequency radio light curve of the sub-pc SMBH binary candidate PG~1302-102}

\author[Basu et al.]
{S. Basu$^{1}$\thanks{E-mail: \href{basusayan2020@gmail.com}{basusayan2020@gmail.com}}, R.P.~Deane$^{1,2}$, K. W. Bannister$^{3}$
\\
$^{1}$ Wits Centre for Astrophysics, School of Physics, University of the Witwatersrand, 1 Jan Smuts Avenue, Johannesburg, 2000, South Africa\\
$^{2}$ Department of Physics, University of Pretoria, Private Bag X20, Pretoria
0028, South Africa \\
$^{3}$ ATNF, CSIRO, Space and Astronomy, PO Box 76, Epping, NSW 1710 Australia
}

\date{Accepted XXX. Received YYY; in original form ZZZ}

\pubyear{\the\year{}}

\begin{document}
\label{firstpage}
\pagerange{\pageref{firstpage}--\pageref{lastpage}}
\maketitle

\begin{abstract}
Supermassive black hole binary (SMBHB) candidates with sub-parsec (sub-pc) orbital separations are rare, high-value astrophysical systems, and they are thought to be the dominant sources that generate the recently detected nanoHertz (nHz) gravitational waves (GWs) background. 
Alongside the search for localized nHz GW sources from SMBHBs using pulsar timing arrays, an increasing number of candidates have been identified from electromagnetic synoptic surveys measuring the optical light curves of large quasar samples over increasingly long time baselines.
The optical light curve of the quasar PG~1302-102 at a redshift of $z=0.278$ exhibits a pronounced, smoothly varying periodic signal of $\sim$$5.2$ years over approximately 20 years. 
Although the cause of this periodicity is still under debate, a plausible explanation is a binary system of two SMBHs with a sub-pc separation. 
At this close distance, the $10^{8.3-9.4} M_{\odot}$ nuclear black holes in PG 1302-102 are expected to merge within $\sim$$10^{5}$ yr solely due to GW emission.
We present a multi-frequency radio light curve analysis of PG~1302-102 over a $\sim$19-year baseline with observations between 2005 and 2024 drawn from the Australian Telescope Compact Array (ATCA). 
The ATCA radio light curve is highly variable at 5.5~GHz but does not reveal any significant evidence of the periodicity seen at optical wavelengths. 
However, we report fractionally stronger evidence of periodicity in the in-band spectral index time-series at 5.5~GHz, with a period consistent with the optical periodicity. 
Furthermore, the recently reported change in optical periodicity also seems to be present in the radio in-band spectral index light curve.
These two results suggest that radio and optical periodicity may be physically coupled despite very different emission mechanisms.
We argue that this seemingly linked radio-optical coupling may reinforce the SMBHB interpretation, despite the optical periodicity change, and provides an intriguing case study ahead of the LSST-SKAO era.

\end{abstract}


%%%%%%%%%%%%%%%%%%%%%%%%%%%%%%%%%%%%%%%%%%%%%%%%%%

%%%%%%%%%%%%%%%%% BODY OF PAPER %%%%%%%%%%%%%%%%%%

\section{Introduction} \label{introduction}
Supermassive black holes (SMBHs) are astrophysical objects residing in the nuclear regions of galaxies and are believed to play a fundamental role in galaxy evolution \citep{2013ARA&A..51..511K}.
During galaxies mergers, the central black holes are expected to undergo dynamical interactions that lead to the formation of supermassive black hole binaries (SMBHBs; \citealp{1980Natur.287..307B}) with separations of order of $\lesssim$10 parsecs (pc). 
The pair of active SMBHs, at a scale of approximately a few kiloparsecs (kpc), may be observed as dual active galactic nuclei (AGN; \citealp{2003ApJ...582L..15K, 2008MNRAS.386..105B, 2011ApJ...740L..44F, 2014MNRAS.437...32W, 2015ApJ...806..219C, 2018Natur.563..214K, 2024MNRAS.531L..76Z, 2025ApJ...986..101L, 2026ApJS..283...70D}).
As the separation continues to decrease by ejecting stars via three-body interactions during in-spiral phase, the pair becomes a gravitationally bound SMBHB at a sub-pc separation before finally merging to form a single black hole \citep[e.g.,][]{2000MNRAS.311..576K, 2016A&A...588A.125R, 2019A&ARv..27....5B, 2023arXiv231016896D} emitting gravitational waves (GWs) (see \citealp{2025CQGra..42q5016B} for a review). 
The binary is predicted to spend most of its lifetime at separations ranging from 0.01 to 1 pc in the GW-dominated zone \citep{1980Natur.287..307B}. 
The phase immediately preceding the merger (the so-called in-spiral) is of particular interest, because it is predicted to be the period when the system emits nanoHertz (nHz) to microHertz ($\mu$Hz) GWs, which have been actively sought through pulsar timing arrays (PTAs; \citealp{2009MNRAS.394.2255S, 2013CQGra..30v4010M}) over the past decade and now \citep{2024ApJ...963..144A, 2024PhRvD.109j1301B, 2024PhRvD.109j1304L, 2024PhRvD.109j3012J, 2024A&A...689A.107L}.
Identifying sub-pc SMBHB candidates is therefore important for constraining gravitational waveform models, refining merger rate predictions, and improving the accuracy of multi-messenger observations.

However, identifying sub-pc SMBHBs through electromagnetic means has proven to be challenging (see \citealp{2019NewAR..8601525D, 2023arXiv231016896D} for reviews). 
Directly imaging such close binaries is feasible only for sources within $\approx$$100$ Mpc through very long baseline radio interferometry (VLBI) observations at~GHz frequencies. 
The closest binary to date has been confidently imaged directly, discovered in the radio galaxy 0402+379, and has a projected separation of 7.3~pc \citep{2006ApJ...646...49R}, corresponding to an orbital period of $\sim$$10^{4}$ years \citep{2017ApJ...843...14B}. 
Additionally, there is a 0.35 pc binary candidate detected in the Seyfert galaxy NGC 7674 at a distance of 116 Mpc \citep{2017NatAs...1..727K}.
Such systems may represent the late stages of galaxy mergers, consistent with the findings of \cite{2018Natur.563..214K}, who show that luminous, obscured accreting black holes are frequently associated with merging galaxies. 
Recent radio studies have also highlighted jet precession as a potential observational signature of SMBHBs.
A substantial number of AGN exhibit curved or evolving radio jet morphologies that have been interpreted in terms of binary-driven precession or orbital motion, although alternative explanations such as accretion disk instabilities and Lense–Thirring precession may also contribute.
Prominent SMBHB candidates discussed in this context include OJ~287 \citep{2018MNRAS.478.3199B, 2023ApJ...951..106B}, 3C~66B \citep{2003Sci...300.1263S}, S5~1928+738 \citep{2014MNRAS.445.1370K}, and BL Lacertae \citep{2003MNRAS.341..405S}, all of which have shown evidence of jet precession or long-term jet position-angle evolution in VLBI observations.

A promising new approach for identifying potential close SMBHB candidates has been developed within the expanding field of time-domain astronomy, which analyzes the periodic light curves of AGN. 
However, periodicity searches in AGN light curves are complicated by stochastic red-noise variability, which can generate apparent periodic or quasi-periodic features, particularly when only a limited number of cycles are observed (e.g., \citealp{2005A&A...431..391V, 2005MNRAS.362..235V, 2014MNRAS.445..437M}).
Consequently, low-significance peaks in periodograms should be interpreted cautiously, as they may arise from correlated stochastic variability rather than from a true periodic process.

The periodic variation in the light curve has been interpreted to be associated with the SMBH orbital motion \citep[e.g.,][]{2013MNRAS.436.2997D, 2014ApJ...783..134F, 2014PhRvD..89f4060G}. 
The periodicity observed in these light curves could result from fluctuations in the accretion rate onto the black holes (\citealp{graham2015}, hereafter G15) and from the relativistic Doppler boost caused by the motion of gas in the mini accretion disk around the smaller black hole in the binary system \citep{2015Natur.525..351D}. 
This observational signature in light curves has prompted numerous systematic endeavors to search for periodically changing quasars within wide-area synoptic surveys \citep{2015MNRAS.453.1562G, 2015ApJ...803L..16L, 2016ApJ...833....6L, 2016MNRAS.463.2145C, 2016ApJ...827...56Z, 2020MNRAS.499.2245C, 2021MNRAS.500.4025L, 2024A&A...683A.248C}.

G15 proposed the quasar PG~1302-102 as a SMBHB candidate based on observed periodicity of $P_{\rm opt}$=1884$\pm$88 days measured in its optical V-band light curve over a $\sim$9 yr baseline. 
PG 1302-102 is a bright, flat-spectrum radio quasar (FSRQ) at a redshift $z=0.278$ \citep{1996ApJS..104...37M} hosted by an elliptical galaxy as is typical for radio-loud quasars. 
The host galaxy of PG~1302-102, as observed in $H-$ and $R-$band imaging, is classified with asymmetric and tidal tail morphologies, respectively \citep{2006ApJS..166...89G, 2015ApJ...804...34H}. 
These galaxy-scale features suggest that the system has undergone merger activity.
The periodicity was detected in the optical light curve from the Catalina Real-Time Transient Survey (CRTS; \citealp{2009ApJ...696..870D}). 
The light curve also included data from the Lincoln Near-Earth Asteroid Research (LINEAR; \citealp{2011AJ....142..190S}) survey, extending $\sim$$0.5$-cycle before the CRTS data. 
G15 interpreted the periodicity as a consequence of the orbital motion of a SMBHB. 
The broad emission lines observed in the spectrum of PG~1302-102 suggest a total binary mass ($M_{BHB}$) to be within the range of $10^{8.3-9.4}M_{\odot}$ \citep{graham2015}. 
G15 also constrain the separation ($r$) between two BHs in the binary system to be $\approx$$0.01$ pc. 
In that case, the post-Newtonian parameter $\epsilon = Gmr^{-1}c^{-2} \approx 0.002$ ($G$ is the gravitational constant) suggests that the binary already entered the in-spiral GW dominant phase of the merger \citep{2015MNRAS.454.1290K}. 

G15 outlined three possibilities that could cause the observed optical variability: (1) precession of the jet causing the change in inclination angle that causes a change in Doppler boosting; (2) a temporary hot spot in the inner region of the accretion disk; or (3) a warped accretion disc might emit light in a way that is variable over time. 
Since different parts of the disc may be tilted toward or away from the observer at different times, the observed brightness can fluctuate periodically. Additionally, as the warp changes, the inclination angle of emission regions can change, affecting Doppler boosting and relativistic beaming.

\cite{2015Natur.525..351D}, using hydrodynamic simulations of circumbinary disks in an unequal-mass (black hole mass ratio $q=M_{1}/M_{2}\leq0.1$) SMBHB system, found that the circumsecondary disk dominates both the total flux and variability, with relativistic beaming explaining the optical periodicity. 
The UV-boosting model for PG~1302-102 explains its periodic variability as relativistic Doppler boosting from the binary system, where emission from an orbiting accretion disk is amplified as it moves toward the observer.
\cite{2015MNRAS.454.1290K} investigated the pc-scale and kpc-scale radio properties of PG~1302-102 jet to explore jet variability, morphological characteristics, gravitational wave emission during the inspiral, and associated timescales, also providing a constraint of $q\geq0.08$. 
At the kpc scale, they inferred a two-sided, asymmetric jet with a wider intrinsic half-opening angle and inclination, potentially exhibiting an expanding helical structure.

While multi-wavelength analyses of PG~1302-102 in the UV \citep{2015Natur.525..351D}, Infrared (IR) \citep{2015ApJ...814L..12J} and radio \citep{2015MNRAS.454.1290K, 2016MNRAS.463.1812M} favour the SMBHB model, recent optical light curve analysis by \cite{2018ApJ...859L..12L} argues against the SMBHB interpretation. 
\cite{2018ApJ...859L..12L}, adding more recent optical light curve sampling observations (2012.1 to 2018.1) from the All-sky Automated Survey for Supernovae (ASAN-SN; \citealp{2014ApJ...788...48S, 2017PASP..129j4502K}), showed the best-fit period of the LINEAR+CRTS+ASAS-SN light curve to be $2012$$\pm$$250$ days which indicates an increase in periodicity reported by G15. 
Recent studies by \cite{2024ApJ...964..167L} and \cite{2025A&A...693A.117R} have investigated the nature of PG~1302-102 through multi-wavelength observations and spectral analysis. 
\cite{2024ApJ...964..167L} conducted intensive broadband reverberation mapping (IBRM) using the Swift and the Las Cumbres Observatory Global Telescope (LCO) observations, finding that the observed UV/optical lag spectrum follows the expected $\tau\varpropto\lambda^{4/3}$ relation for a standard disk, with no strong evidence supporting the presence of a second BH. 
The measured disk size is twice as large than expected for a minidisk around a secondary BH \citep{2015Natur.525..351D} in a binary system. 
Similarly, \cite{2025A&A...693A.117R} analyzed spectroscopic features using the Very Large Telescope (VLT) observations.
In a close SMBHB scenario, one may expect signatures such as double-peaked or velocity-shifted broad emission lines arising from distinct or dynamically modulated broad-line regions associated with each black hole. 
However, no clear evidence for such kinematic signatures was found, which further disfavors a binary interpretation of the system.
These recent findings argue that PG~1302–102 is more consistent with a single SMBH system undergoing disk-driven variability rather than a binary scenario. 
However, the presence of periodic variability in multiple wavebands, including radio and optical, raises the possibility of alternative mechanisms which merit further monitoring and investigation.

Motivated by the possibility of an independent measure of the periodicity, we analyzed archival radio observations from the Australia Telescope Compact Array\footnote{\url{https://www.narrabri.atnf.csiro.au}} (ATCA).
Our paper is organized as follows: in Section \ref{obs}, we outline the observations used for the analysis and present the results.
This section showcases the radio periodicities calculated using the ATCA flux density and in-band spectral index.
In Section \ref{discussion}, we discuss the findings from the analysis and how they complement the SMBHB models proposed in previously published works.  
Finally, we draw our conclusions in Section \ref{conclusion}. 
Throughout this paper, we assume the Planck 2018 cosmology \citep{Planck2018} and define the spectral index as $\rm S_{\nu} \propto \nu^{\alpha}$.

\section{Observations \& Results} \label{obs} 
\subsection{ATCA data}
We obtained flux measurements of PG 1302-102 at 2.1, 5.5, 17, and 33 GHz from the ATCA calibrator database\footnote{\url{https://www.narrabri.atnf.csiro.au/calibrators/calibrator_database.html}} (version 3).
Visibility data were calibrated and reduced in \textsc{miriad} package \citep{miriad} using the observatory-developed automated pipeline. 
The observations were obtained using the standard \textsc{CABB} continuum setup, providing an instantaneous bandwidth of approximately 2~GHz.
Bandpass and flux density calibration were done using observations of the standard ATCA flux density calibrator B1964-638. 
The calibration procedure includes flagging bad data, correcting for instrumental effects, and compensating for atmospheric opacity. 
The source flux density is measured by fitting models to the visibility data.  
Each flux density is usually accompanied by an uncertainty estimated from the root mean square (RMS) value of the visibility amplitudes after the best fit is subtracted. 

\begin{figure}
  \centering
  \includegraphics[width=\linewidth]{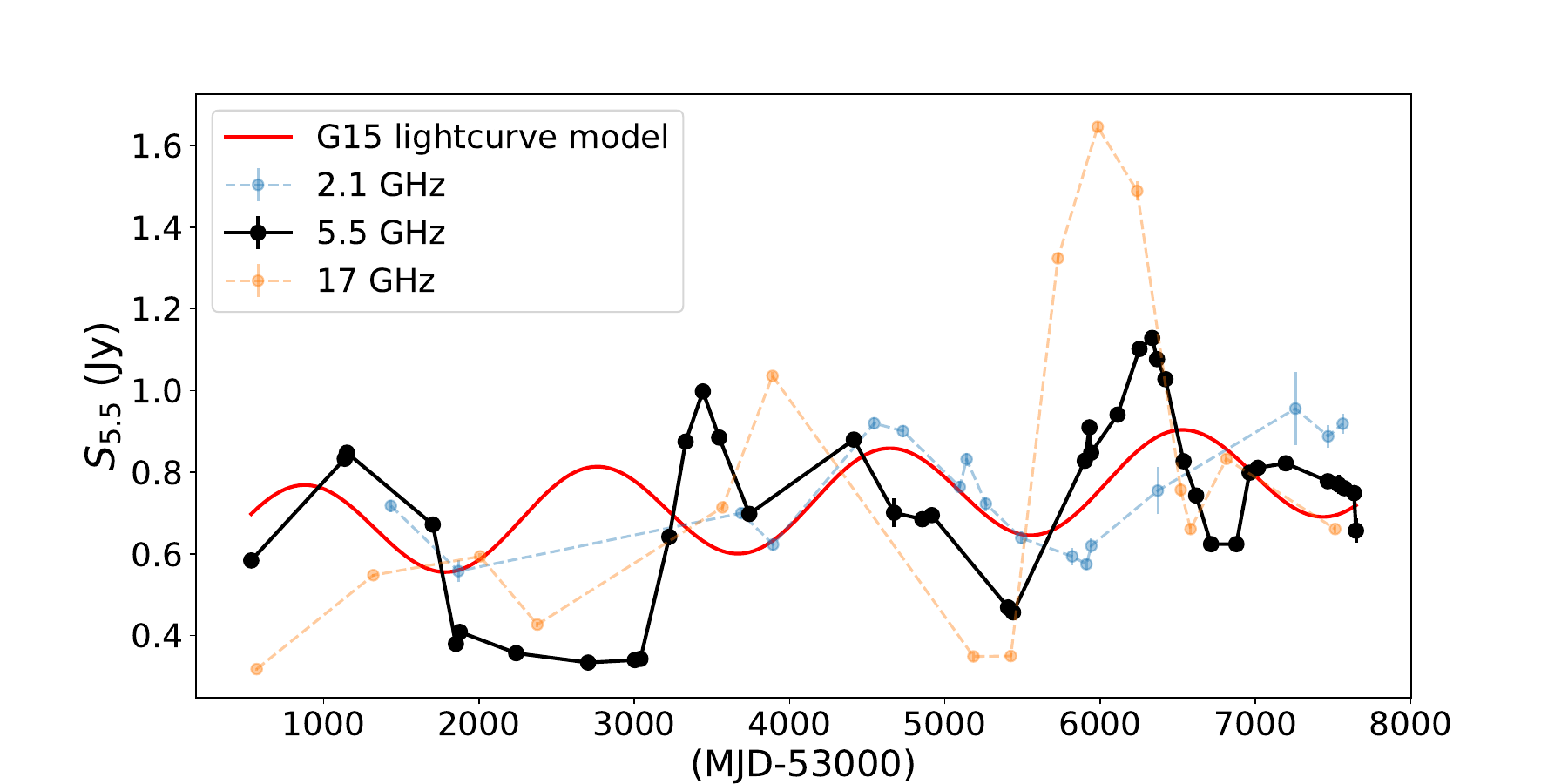} 
  \caption{Multi-frequency ATCA radio light curves of PG 1302-102. The red line is the sinusoid fit with a period of 1884 days and phase reported by G15 using the CRTS optical observations.}
  \label{fig:atca flux}
\end{figure}

Table \ref{table:obs} shows detailed information about observations at each frequency. 
\begin{table}
\centering
\begin{tabular}{|c|c|c|c|}
\hline
Frequency (MHz) & No. of obs. & Timespan (Year) & Avg. cadence (Days)  \\
\hline
2100 & 17 & 2007.9 - 2024.6 & 383 \\
\hline
5500 & 41 & 2005.4 - 2024.9 & 177 \\
\hline
17000 & 15 & 2005.5 - 2024.5 & 496 \\
\hline
33000 & 12 & 2008.4 - 2023.2 & 491 \\
\hline
\end{tabular}
\caption{Statistics of the ATCA observations at each frequency between MJD 53534 and MJD 60647.}
\label{table:obs}
\end{table}
For the periodicity analysis, we selected flux measurements at 5.5 GHz ($S_{5.5}$) as it had the most number of flux measurements among all the frequencies and the highest sampling density.
In Figure \ref{fig:atca flux}, we show the $S_{5.5}$ radio light curve over 19 years (MJD $\approx$ 53,534 to MJD $\approx$ 60,647) from 41 epochs at multiple frequencies, but highlighting the 5.5~GHz observations. 
This results in an average observing cadence of $\sim$6 months, which results in roughly 10 samples per observed-frame period. 
Please note, the 33~GHz data are excluded in the figure because they show unusually large variability and an extreme flare ($\sim$3.4~Jy) that is inconsistent with both the lower-frequency observations and the light-curve model. 
These points are likely affected by larger calibration/systematic uncertainties and would bias the overall fit.

\subsection{Periodicity Searching}
We search for periodic variation in the light curves via Lomb-Scargle periodogram (LSP) \citep{1976Ap&SS..39..447L, 1982ApJ...263..835S}.
This is a useful method for identifying and characterizing periodic variations in irregularly sampled astronomical data, such as time series with gaps caused by observational constraints, eliminating the need for data gap interpolation (e.g., \citealp{2018Galax...6..136B}, \citealp{2018ApJS..236...16V}). 
The LSP is a widely used method in time series analysis, derived from the Discrete Fourier Transform (DFT) method (e.g., \citealp{2018Galax...6..136B}, \citealp{2023A&A...678A.100P}).
Unlike traditional Fourier transforms, the LSP uses a least-squares fit of sinusoidal models, adjusting both frequency and phase to minimize residuals, making it particularly effective for identifying quasi-periodic oscillations (QPOs) in noisy, irregularly sampled data. 
Its normalization facilitates the comparison of peaks, while the inclusion of phase information reduces aliasing and spurious signals.  
Nonetheless, despite these advantages the model fundamentally assumes periodicity, which may not necessarily be present. 
In the case of the ATCA light curves, the limited temporal baseline and small number of apparent cycles make it difficult to robustly distinguish a genuine periodic component from stochastic red-noise fluctuations. 
Therefore, the observed peaks should not be interpreted as standalone evidence for periodicity.

In addition to the single-frequency radio light curves, we also obtained the 5.5 GHz in-band spectral index light curve ($\alpha_{5.5}$) from the database.
In the database, the spectral index is defined through the relation ($S \propto \nu^{\alpha}$) and is derived from the frequency-dependent model fitted to the calibrator flux densities. 
The reported value corresponds to the local spectral slope, ($d\log S/d\log\nu$), evaluated at the reference frequency. 
For the logarithmic polynomial model commonly adopted in the database, ($\log S = a + b\log\nu + c(\log\nu)^2$), giving ($\alpha = b + 2c\log\nu$). 
Further details of the spectral-index calculation are available in the ATCA calibrator database documentation (see \href{https://www.narrabri.atnf.csiro.au/calibrators/calibrator_database_documentation.html#interpreting-spectral-indices}{ATCA Calibrator Database Documentation}).
Please note that due to very few observations, periodicity at other frequencies has not been calculated.

To analyse the periodic variation, we divide the $S_{5.5}$ observations into two parts: (1) pre-G15 and (2) post-G15. 
The pre-G15 segment includes observations between 2005 and 2014, and the  post-G15 segment includes observations between 2016 and 2024. 
The motivation behind this segmentation is to analyse the periodic nature of PG~1302-102 in the radio frequency during the epoch when G15 reported the $\sim$5.2-yr periodicity and to investigate how this periodicity evolves in subsequent epochs.  
We also divide the $\alpha_{5.5}$ timeseries into two parts in same way it was done for the $S_{5.5}$.
It is worthy to note that the light curves are sparsely and irregularly sampled, which introduces significant spectral leakage and window-function effects in the periodogram analysis. 
These effects can enhance low-frequency power and complicate the interpretation of red-noise significance levels.

\begin{figure}
  \centering
  \includegraphics[width=\linewidth]{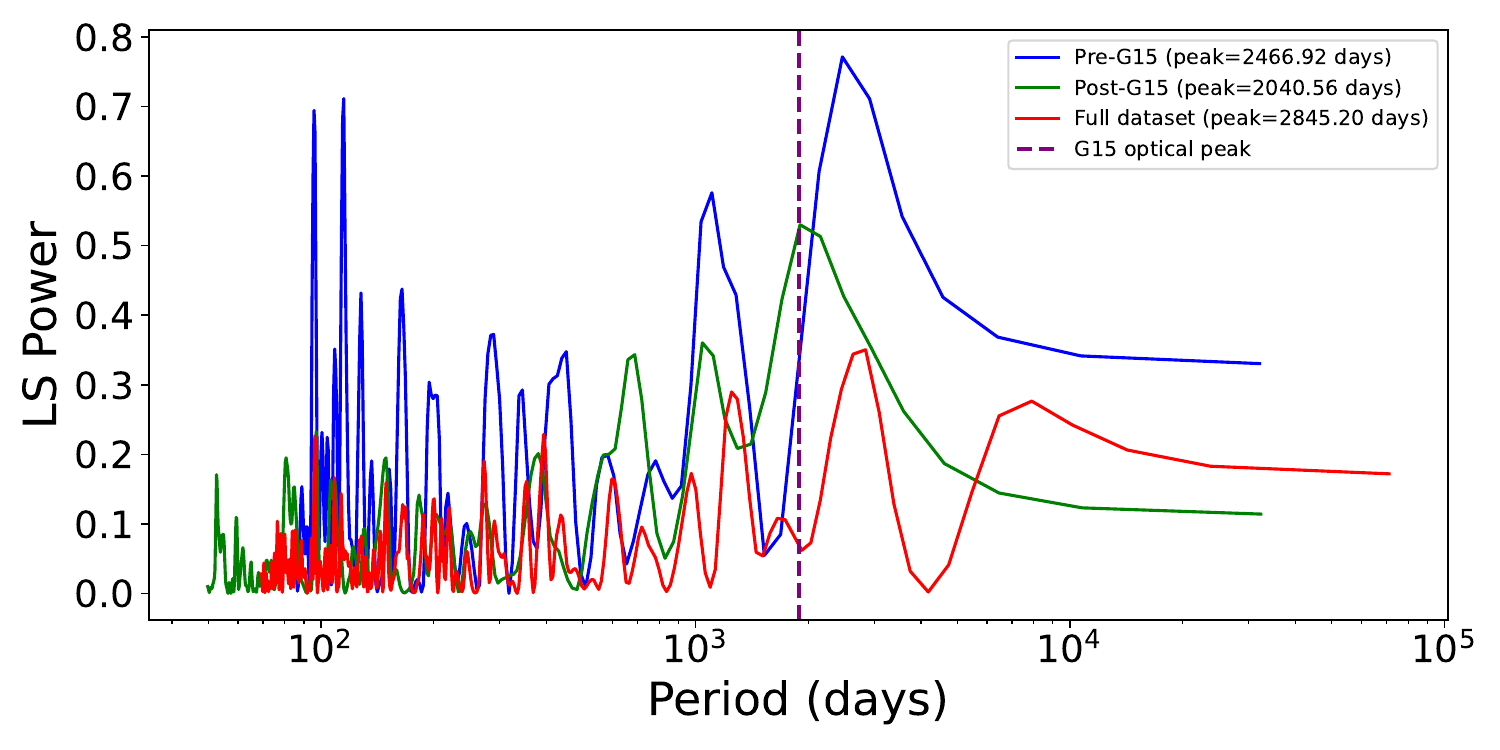} 
  \caption{Lomb-Scargle periodogram of ATCA 5.5 GHz flux density ($S_{5.5}$) light curve. The plot contains periodograms from pre-G15, post-G15 epoch, as well as all the epochs. The vertical line represents G15 optical peak with $1884$ days.}
  \label{fig:atca LS}
\end{figure} 

\begin{figure}
  \centering
  \includegraphics[width=\linewidth]{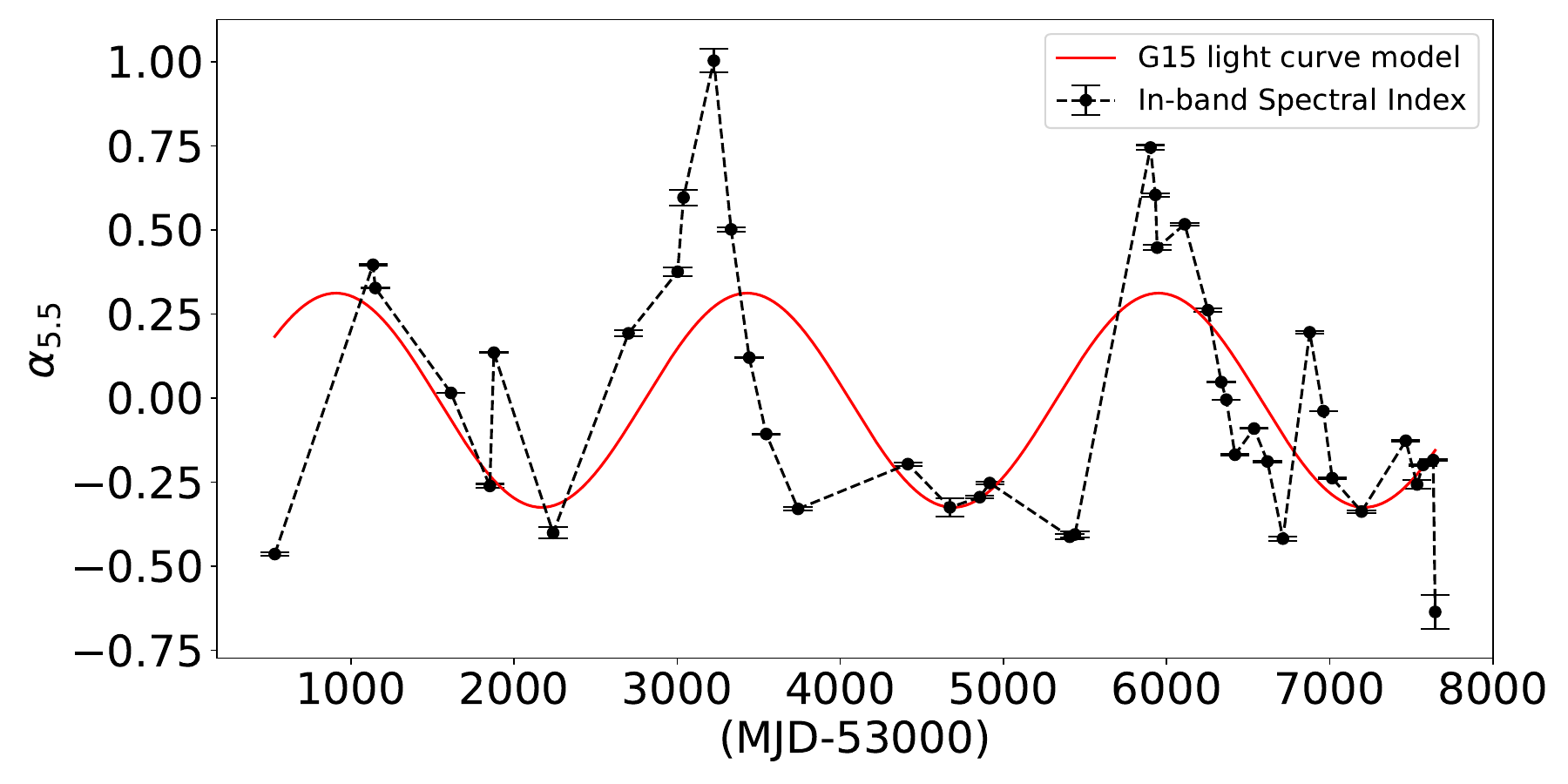} 
  \caption{ATCA 5.5~GHz in-band spectral index ($\alpha_{5.5}$) time series between 2005-2024. The red line is the sinusoid with a period of 1884 days reported by G15.}
  \label{fig:atca spectral}
\end{figure}
\begin{figure*}
  \centering
  \includegraphics[width=\linewidth]{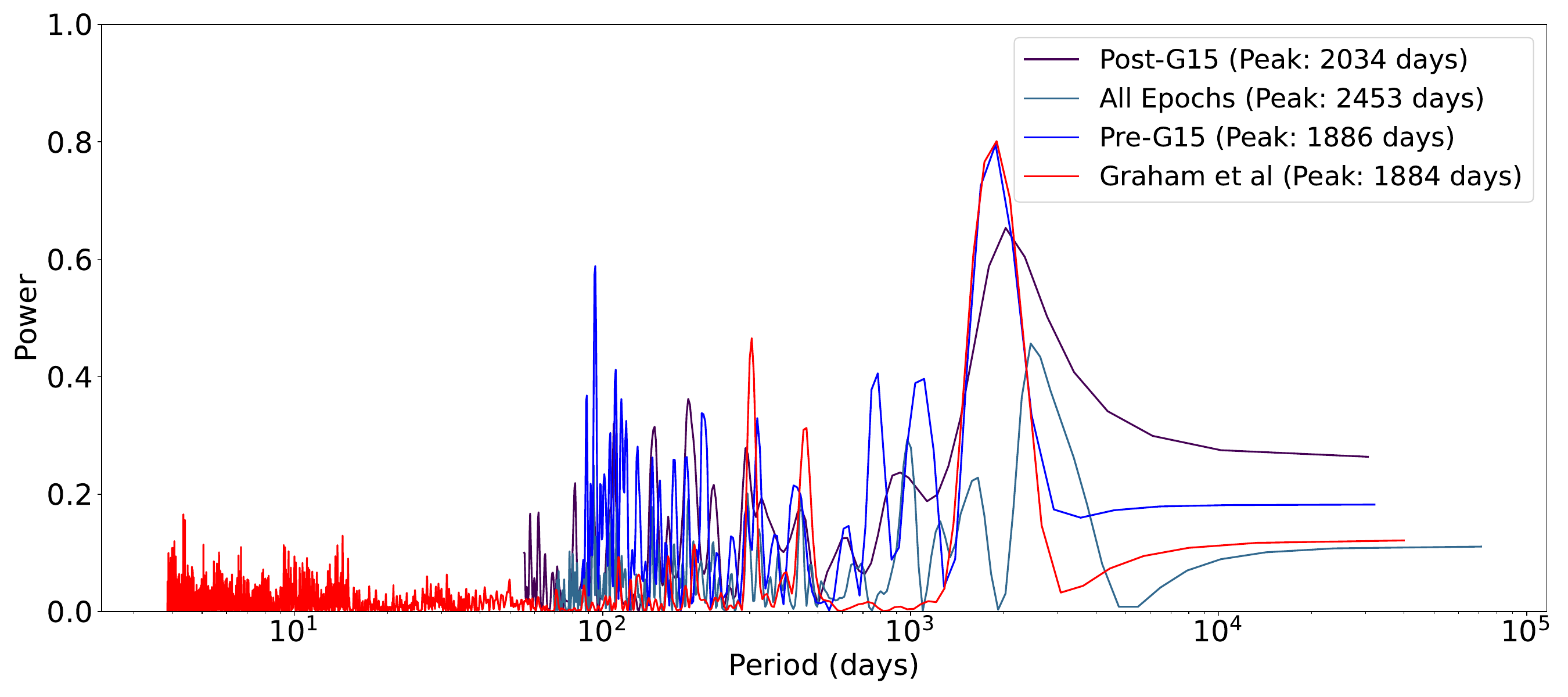} 
  \caption{Lomb-Scargle periodogram of ATCA 5.5~GHz in-band spectral index ($\alpha_{5.5}$). The plot contains periodograms from the pre-G15, and post-G15 epochs as well as all the epochs. In this plot, we also include a periodogram from the ASAS-SN observations that \protect\cite{2018ApJ...859L..12L} used. Note that the observations \protect\cite{2018ApJ...859L..12L} used were from $\sim$two years and covers only the half of the $\sim$$5.2$ years periodicity reported by G15.}
  \label{fig:atca spectral LSP}
\end{figure*} 
   
\begin{figure}
  \centering
  \includegraphics[width=\linewidth]{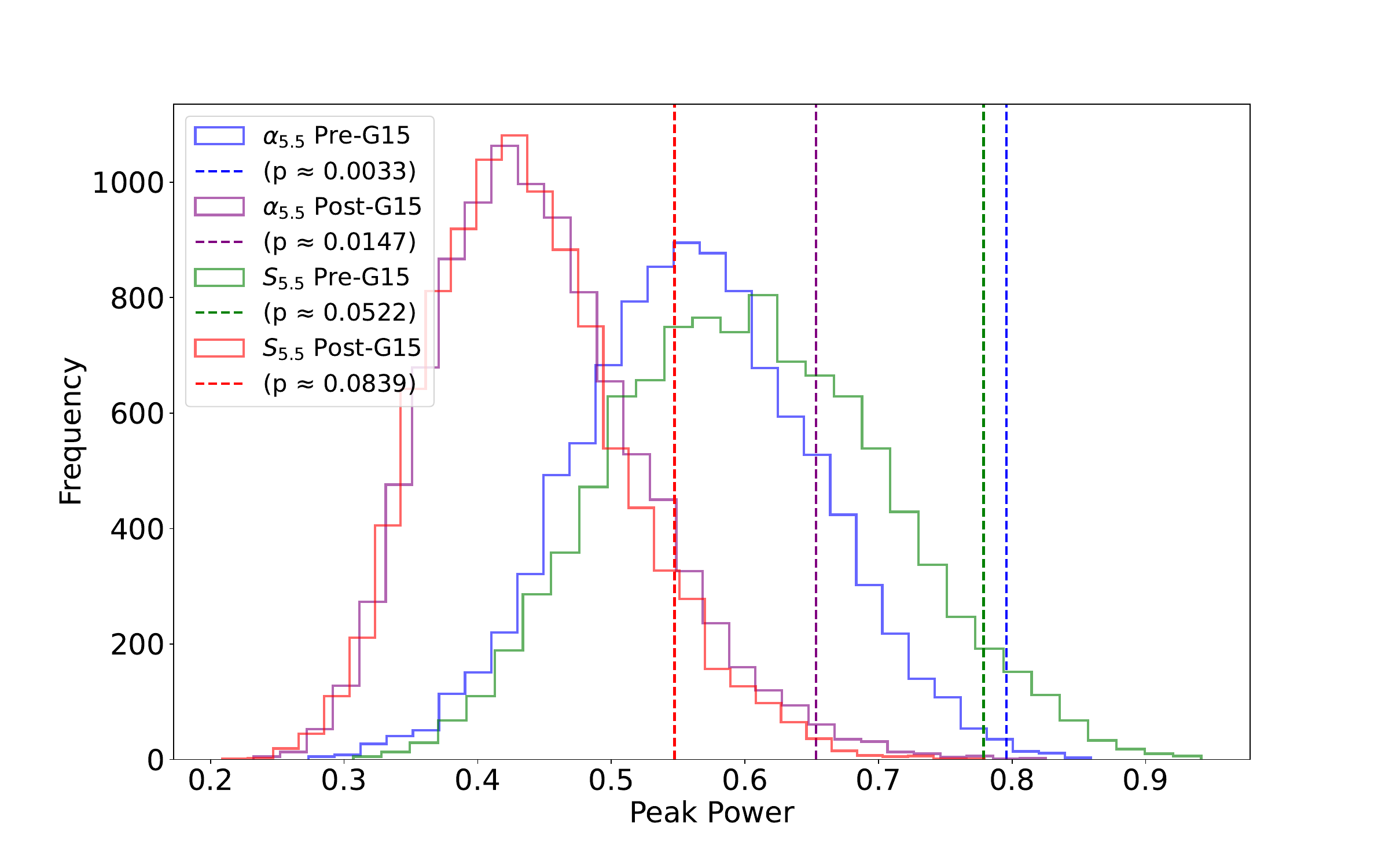} 
  \caption{Peak power distribution with the p-value of ATCA 5.5~GHz in-band spectral index ($\alpha_{5.5}$) and flux density ($\rm S_{5.5}$) in the pre- and post-G15 analysis.}
  \label{fig:peak_power}
\end{figure}

\begin{itemize}
        \item {\bf Pre-G15: } In Figure \ref{fig:atca LS} and \ref{fig:atca spectral LSP}, we plot results from the periodicity analysis between epochs from 2005-2014 (labeled as pre-G15 in the figure) for $S_{5.5}$ and $\alpha_{5.5}$.  
        We detect significant periodicity ($P_{\alpha}$) $1886$$\pm221$ days for $\alpha_{5.5}$, which is consistent with the periodicity reported by G15. 
        However, we do not find any significant periodicity in $S_{5.5}$ light curve and the periodicity ($P_{s}$) is $2466$$\pm$$379$ days. 
        The uncertainty in the periodicity is estimated via a Monte Carlo simulation comprising $10^4$ realisations, in which the spectral index values are resampled within their statistical uncertainties while preserving their associated MJDs.
        We calculated the uncertainty as the half-width half-maximum of the spectral peak in the LSP, assuming it is approximated by a Gaussian profile.  
        To assess the statistical significance of the detected period, we compute the False Alarm Probability (FAP), a standard metric used to evaluate the likelihood that a peak arises from stochastic variability in a time series. 
        In the white-noise case, each Monte Carlo realisation is constructed by perturbing the spectral index measurements within their observational uncertainties, while keeping MJDs fixed; thus, each realisation represents a statistically plausible noise realisation under the assumption of independent Gaussian errors.
        
        However, given the presence of correlated variability in astrophysical time series, we also consider a red-noise model in which synthetic light curves are generated from a power-law power spectral density $P(f) \propto f^{-\beta}$ consistent with the observed stochastic variability. For each realisation, we compute the periodogram and record the maximum power. The FAP is then defined as the fraction of simulations in which the maximum periodogram power exceeds that of the observed data.

        If only a small fraction of realisations exceed the observed peak power, the null hypothesis of purely stochastic variability can be rejected with the corresponding confidence level.
        
        We report that the confidence level in the pre-G15 periodicity to be $98.54$ per cent for the $\alpha_{5.5}$ and $97.7$ per cent for the $S_{5.5}$. 
        To assess the statistical significance of the periodicity detected in the time-series data, we also calculate the p-value associated with the peak power in the LSP. 
        The p-value represents the probability of observing a peak power as high as, or higher than, the detected value under the null hypothesis that the data contains no true periodicity and the observed variations are due to random noise. 
        This was determined by generating a distribution of peak power values from Monte Carlo realisations, where the $\alpha_{5.5}$ and $S_{5.5}$ were randomly permuted while preserving the original observation times (MJD). 
        The peak power corresponds to the maximum value in the periodogram for each realization, allowing us to assess the statistical significance of the detected periodicity.
        The p-value quantifies the likelihood that the detected signal is due to chance, with smaller values indicating higher confidence in the periodic signal.
        The p-value for $\alpha_{5.5}$ is $3.3\times10^{-3}$ and for $S_{5.5}$ is $5.2\times10^{-2}$. 
        In Figure \ref{fig:peak_power} we show the distribution of peak power and p-value of $\alpha_{5.5}$ and $S_{5.5}$.  
        To assess the significance of periodic signals, we modelled the intrinsic variability as a power-law red-noise process with spectral index $\beta$=0.13, estimated from the observed periodogram.
        As mentioned earlier, the periodogram exhibits a dominant peak at $1886\pm221$ days. 
        However, comparison with the simulated red-noise distribution shows that this feature does not exceed the $99$ per cent FAP threshold (Table \ref{tab:fap_results}).
\end{itemize}
\begin{itemize}
\item{\bf Post-G15: } In Figure \ref{fig:atca LS} and \ref{fig:atca spectral LSP}, we plot the periodograms between 2016-2024 (labelled as Post-G15 in the figures). 
We calculate $P_{\alpha}$ to be  $2034\pm260$ days and $P_{s}$ to be $2040\pm271$ days. 
        The uncertainty is calculated in the same way as it was done for Pre-G15.
        We calculate the confidence level in the post-G15 periodicity to be $99.03$ per cent for $\alpha_{5.5}$ and $90.6$ per cent for $S_{5.5}$.
        The p-values are $1.47\times10^{-2}$ and $8.39\times10^{-2}$ for $\alpha_{5.5}$ and $S_{5.5}$ respectively.
        To evaluate the significance of candidate periodicities, we modelled the intrinsic variability as a power-law red-noise process with $\beta$ = 0.67, estimated from the observed periodogram.
        The comparison with the red-noise simulations shows that this feature does not exceed the $99$ per cent false alarm probability threshold (Table \ref{tab:fap_results}). The peak is therefore consistent with stochastic red-noise variability and does not constitute a statistically significant periodic detection in the post-G15 segment.
\end{itemize}
\begin{itemize}
\item{\bf All epochs: } We show the periodograms (labelled as All Epoch in Figure \ref{fig:atca LS} and \ref{fig:atca spectral LSP}) as well as the timeseries (Figure \ref{fig:atca spectral}) where we combine all available ATCA observations between 2005-2024. 
We calculate the periodicity to be $2390\pm164$ days and the confidence level to be $97.23$ per cent for $\alpha_{5.5}$ and $2592\pm207$ days and $99.7$ per cent for $S_{5.5}$. 
The full time series was analysed combined with the red-noise model. 
The best-fit red-noise index $\beta$=0.42, indicating weakly correlated variability in the spectral index time series. 
Monte Carlo simulations of red-noise light curves yield a 99 per cent false alarm probability threshold that is slightly below the observed peak power; however, the peak does not exceed the 99.9 per cent significance level (Table \ref{tab:fap_results}).
\end{itemize}
Notably, the $\alpha_{5.5}$ appears to provide more robust measure of periodicity than $S_{5.5}$, as it is less affected by intrinsic source variability and external propagation effects, making it a more reliable tracer of underlying physical processes. 
These results suggest that the periodic variability observed in the radio band is consistent with the previously reported optical periodicity.
\begin{table*}
\centering
\begin{tabular}{|c|cc|cc|cc|cc}
  \hline
  \multirow{2}{*}{Epoch} & \multicolumn{2}{c|}{$S_{5.5}$} & \multicolumn{2}{c|}{$\alpha_{5.5}$} & \multicolumn{2}{c|}{LINEAR+CRTS} &  \multicolumn{2}{c|}{LINEAR+CRTS+ASAS-SN} \\
  \cline{2-9}
   & Period ($\rm P_{s}$) & N-sigma & Period ($\rm P_{\alpha}$) & N-sigma & Period ($\rm P_{opt}$) & N-sigma & Period & N-sigma \\
   & (days) & (\%) & (days) & (\%) & (days) & (\%) & (days) & (\%)\\
  \hline
  2005-2024 & $2592\pm207$ & 99.7 & $2390\pm164$ & 97.2 & .. & .. & .. & .. \\
  2005-2014 & $2466\pm379$ & 97.7 & $1886\pm221$ & 98.5 & .. & .. & ..  & .. \\
  1995-2015 & .. & .. & .. & .. & $1884\pm84$ & .. & .. & .. \\
  1995-2017 & .. & .. & .. & .. & .. & .. & $2012\pm250$ & .. \\
  2016-2024 & $2040\pm271$ & 90.6 & $2034\pm260$ & 99 & .. & .. & .. & .. \\
  \hline
\end{tabular}
\caption{Periodicity and confidence on the period derived from the LSPs for ATCA observations, LINEAR+CRTS (by G15) and LINEAR+CRTS+ASAS-SN \citep{2018ApJ...859L..12L}.}
\label{table:periodicity table}
\end{table*}

\begin{table*}
\centering
\begin{tabular}{lcccccc}
\hline
Segment & No. of Epochs & $\beta$ & $P_{\mathrm{peak}}$ (days) & Peak Power & 99\% FAP Threshold & Significance \\
\hline
Pre-G15   & 15 & 0.13 & 1886 & 0.80 & 0.82 & Not significant \\
Post-G15  & 26 & 0.67 & 2034 & 0.65 & 0.65 & Not significant \\
All Epochs & 41 & 0.42 & 2476 & 0.45 & 0.440 (99\%) / 0.502 (99.9\%) & Marginal (99--99.9\%) \\
\hline
\end{tabular}
\caption{Red-noise False Alarm Probability (FAP) analysis of the spectral index variability across different temporal segments. The red-noise model is described by a power-law power spectral density. Significance levels are derived from Monte Carlo simulations preserving the observational sampling.}
\label{tab:fap_results}
\end{table*}

\section{Discussion} \label{discussion}
G15 proposed a Doppler boosting model for PG 1302–102 based on their discovery of a significant 1884-day periodicity in its optical light curve from the CRTS V-band observations. 
The periodicity was confirmed by UV observations from the Galaxy Evolution Explorer (GALEX; \citealp{2005ApJ...619L...1M}) survey and was attributed to Doppler boosting of emission from the accretion disk surrounding the more luminous black hole in the binary, modulated by the orbital motion. 
The UV data, which probe the hotter regions of the accretion disk, revealed a periodic signal with a noticeable amplitude boost at shorter wavelengths, reinforcing this interpretation. 
This periodic signal was interpreted as the binary’s orbital period, corresponding to a separation of $\sim$$0.01$ pc. 
At this scale, relativistic effects such as gravitational wave emission and strong dynamical interactions become significant.

Subsequent studies by \cite{2015MNRAS.454.1290K}, \cite{2016MNRAS.463.1812M}, and \cite{2018A&A...615A.123Q} have expanded the understanding of PG 1302-102’s variability by incorporating high-resolution radio observations.
The periodic variation can be interpreted using models that incorporate jet precession (e.g., \citealp{2000A&A...360...57R, 2003MNRAS.341..405S, 2013MNRAS.428..280C}) or helical jet structures (e.g., \citealp{1993ApJ...411...89C, 1998ApJ...500..810T, 2015ApJ...805...91M}).
\cite{2015MNRAS.454.1290K} analyzed archival VLBA and VLA radio data to reveal the pc-scale and kpc-scale structure of the radio jet in PG 1302-102.
Their results showed that the jet has a half-opening angle of $0.20^{\circ}$$\pm$$0.05^{\circ}$, determined by the orbital motion of the black hole emitting the jet. 
The distinct inclinations of the kpc-scale ($\sim$$65.5^{\circ}$) and pc-scale ($\sim$$2.2^{\circ}$) jets in PG 1302-102 challenge the idea that the pc-scale jet is precessing within the kpc-scale jet’s cone. 
A simple precession model cannot explain this misalignment, suggesting more complex dynamics. 
One possibility is a past spin-flip event, though this is unlikely based on system parameters. 
Alternatively, the two jets may have originated from different black holes in a binary system, with one jet turning off while the other became active. 
An alternative scenario outlined by \cite{2015MNRAS.454.1290K} involves a triple black hole interaction, where a past merger reoriented the spin and altered the jet direction. 
However, the long precession period of the kpc-scale jet suggests a gravitational lifetime of millions of years, making a recent final merger unlikely. 
These uncertainties highlight the need for further investigation into the jet formation mechanisms in SMBHB systems.
\cite{2016MNRAS.463.1812M} extended the study of PG~1302-102’s jet dynamics by incorporating 2–8 GHz and 15 GHz VLBA radio observations to constrain its pc-scale jet kinematics and spectral properties. They derived a proper motion of $\mu=0.27\pm0.04$ mas $\rm yr^{-1}$, a bulk Lorentz factor $\Gamma \geq 5.1\pm0.8$, and an inclination angle $i\leq11^{\circ}.4\pm1^{\circ}.7$, projected position angle $31^{\circ}.8\pm0.2$, along with a mean spectral index $\alpha=0.31$ (2-8~GHz). 
These parameters characterize a highly relativistic, beamed jet. \cite{2016MNRAS.463.1812M} applied their general relativistic helical jet model \citep{2015ApJ...805...91M} to the radio data, showing that helical motion in the jet could account for up to $53$ per cent of the observed core flux density variability.

\cite{2018A&A...615A.123Q} further investigated jet precession in PG~1302-102, deriving a precession period of approximately 1883 days from the VLBA data used by \cite{2015MNRAS.454.1290K}. 
They attributed this precession to the orbital motion of the SMBHB system, with periodic ejections of superluminal jet components. 
The close alignment between the jet precession period and the optical variability period suggests a possible connection between jet dynamics and disk variability. 
This correlation implies that periodic optical emission may result from changes in the accretion flow, which modulate the jet orientation and emission, possibly due to the influence of a binary companion. 
\cite{2018A&A...615A.123Q} also highlighted the importance of radio jet precession for understanding the periodic optical flux variations reported by G15.

However, as noted in Section \ref{introduction}, subsequent analyses with additional data have refined the periodicity of PG~1302-102.
\cite{2018ApJ...859L..12L}, combining CRTS data with observations from the ASAS-SN survey, found that the optical periodicity increased to 2012 days.
\cite{2024ApJ...964..167L}, using the Swift and LCO observations, argued against the SMBHB model using reverberation mapping.
The results show that the lag spectrum follows the relation $\tau\varpropto\lambda^{4/3}$, as expected for thermal reprocessing in a standard geometrically thin accretion disk. 
This smooth power-law dependence indicates a single, radially continuous emitting structure, whereas a SMBHB system would generally introduce additional complexities or deviations due to multiple accretion flows and orbital modulation.
The inferred accretion disk size, based on lag measurements ($\tau_{0}\approx13.7\pm2.9$ days), is larger than predictions for a secondary minidisk in an SMBHB by \cite{2015Natur.525..351D}.
\cite{2025A&A...693A.117R} analyzed the spectroscopic and photometric properties of PG~1302–102 to test the SMBHB hypothesis.  
Although, a perturbed single-BH BLR model was favoured, the observed multi-component velocity structure, compact nuclear scale, and persistent periodic variability remain fully consistent with expectations for a sub-pc SMBHB system. 
The characteristic optical periodicity ($\sim$$1884$ days in the observed frame) was reassessed, with indications that it may be an artifact of AGN red noise rather than a true binary signature. 
The study also found that the broad-line velocity shifts were $<$$200$ km $\rm s^{-1}$, significantly lower than expected for an SMBHB with a separation of $\sim$$0.01$ pc. 
Instead, they argue that the spectral features and variability patterns were more consistent with a single SMBH with an accretion disk rather than an SMBHB with Doppler-boosted emission. 
Both the studies however, do not exclude alternative scenarios such as jet precession, suggesting that further multiwavelength observations, particularly at radio wavelengths, could help distinguish between models.

Motivated by these arguements, we analyze the $\alpha_{5.5}$ variability in PG~1302-102 where we identified a periodicity of 1886$\pm$221 days in pre-G15 segment of the light curve, consistent with the optical periodicity reported by G15.
The in-band spectral index is expected to be less sensitive to Doppler boosting, as relativistic beaming primarily affects the flux measurements while preserving the slope of an intrinsic power-law spectrum (e.g., \citealp{1995PASP..107..803U}). 
However, red-noise simulations indicate that the pre-G15 peak does not exceed the $99$ per cent FAP threshold, implying that the observed timescale cannot be regarded as a statistically significant periodic detection. 

With the inclusion of follow-up observations, the dominant timescale shifts to $2034$ days in the post-G15 (2016-2024) segment and further to $2390$ days when the full 2005--2024 dataset is considered.
The all-epoch peak is only marginally significant, lying above the $99$ per cent red-noise threshold but below the $99.9$ per cent level, while the post-G15 peak remains below the $99$ per cent threshold. 
Consequently, no statistically robust periodicity is detected in any temporal segment once correlated stochastic variability is taken into account. 
The measured change in periodic variation in the $\alpha_{5.5}$ follows the trend observed in the optical, where the periodicity increased from 1886 days to 2012 days after adding observations from the ASAS-SN survey (\citealp{2018ApJ...859L..12L}). 
Although this similarity does not constitute evidence for a common periodic driver, it suggests that the spectral-index variability may encode long-timescale processes that evolve with increasing observational baseline.

In a SMBHB system with a projected separation of $\sim$$0.01$~pc, if GW emission is dominant, the separation would decrease over time, leading to a shorter orbital period, not a longer one assuming circularised orbits.
However, the shift to a longer periodicity in both optical and radio bands indicates that the periodic variability in PG~1302-102 is not a fixed feature of the system but evolves over time.
In SMBHB systems, jet precession can arise due to the misalignment between the secondary SMBH's orbital plane and the accretion disk of the central SMBH (e.g., \citealp{2000A&A...355..915A, 2000A&A...360...57R, 2004ApJ...602..625C, 2017ApJ...851L..39C, 2022A&A...665L...3D, 2022ApJ...924...93C}). 
This precession period can be longer than the orbital period, and its impact on the observed periodicity becomes more apparent with longer observational baselines.
While optical emission is primarily governed by processes occurring in the accretion disk close to the black holes, radio emission originates from the relativistic jet, which can extend to much larger scales. 
The longer period in the radio band may reflect the contribution of these larger-scale jet dynamics, which operate on a different timescale compared to the inner-disk processes that dominate the optical emission.

Overall, the red-noise analysis indicates that the variability of $\alpha_{5.5}$ is consistent with correlated stochastic processes and does not provide statistically robust evidence for a persistent periodic signal. Nonetheless, the apparent correspondence between the evolution of radio and optical characteristic timescales, together with the relative stability of $\alpha_{5.5}$ compared to $S_{5.5}$, suggests that the in-band spectral index remains a promising diagnostic for future studies of long-term variability in PG~1302$-$102. Continued monitoring and detailed physical modelling in the LSST--SKAO era will be required to determine whether these evolving timescales reflect jet dynamics, accretion-driven variability, or more complex manifestations of an SMBHB system.

\section{Conclusions} \label{conclusion}
We have presented the first long-term study of the ATCA 5.5 GHz radio flux-density light curve and in-band spectral index ($\alpha_{5.5}$) variability of PG~1302$-$102 using observations spanning nearly two decades (2005--2024). While the 5.5 GHz flux-density light curve does not exhibit any statistically significant periodicity, the $\alpha_{5.5}$ time series reveals dominant long-timescale features at $\sim$$1886$ days, $\sim$$2034$ days, and $\sim$$2476$ days in the pre-G15, post-G15, and full datasets, respectively. The earliest timescale is broadly similar to the optical periodicity reported by G15 and to the timescale inferred from VLBA-based jet-precession studies by \cite{2018A&A...615A.123Q}.

However, red-noise false alarm probability analyses demonstrate that the pre-G15 and post-G15 features do not exceed the $99$ per cent significance threshold, while the all-epoch feature is only marginally significant, lying above the $99$ per cent level but below the $99.9$ per cent threshold. Consequently, we do not find statistically robust evidence for a persistent periodic signal in either the radio flux-density or spectral-index variability of PG~1302$-$102. The variability is instead consistent with correlated stochastic processes, with red-noise indices that differ between epochs, indicating that the source variability is non-stationary over decadal timescales.

Nevertheless, the systematic increase in the characteristic timescale of $\alpha_{5.5}$ with the inclusion of additional observations qualitatively mirrors the evolution reported in the optical band. Although this correspondence does not establish a common physical origin, it suggests that the in-band spectral index may encode long-term processes associated with the dynamical evolution of the system. Such behaviour could arise from jet dynamics, precessional phenomena, accretion-state changes, or more complex manifestations of an SMBHB environment, but the present data do not allow these possibilities to be distinguished.

Our results demonstrate that the in-band spectral index provides a valuable and complementary probe of long-term variability in active galactic nuclei and motivate continued multiwavelength monitoring of PG~1302$-$102. Future observations with next-generation facilities, particularly in the LSST--SKAO era and through high-resolution VLBI imaging, will be crucial for determining whether the evolving characteristic timescales represent manifestations of stochastic variability or signatures of underlying physical processes operating in this remarkable system.

\section{Acknowledgements}
We thank Ian Heywood and Sarah Burke-Spolaor for useful inputs in preparation of this paper.
SB acknowledges the South African National Research Foundation (NRF) and SARChI post-doctoral fellowship to conduct this research.  
This paper includes archived data obtained through the ATCA calibrator database. 
The NRF is the intermediary agency between the policies and strategies of the Government of South Africa and South Africa's research institutions. 
We acknowledge the use of the ilifu cloud computing facility, a partnership between the University of Cape Town, the University of the Western Cape, Stellenbosch University, Sol Plaatje University, the Cape Peninsula University of Technology, and the South African Radio Astronomy Observatory. 
The ilifu facility is supported by contributions from the Inter-University Institute for Data Intensive Astronomy (IDIA - a partnership between the University of Cape Town, the University of Pretoria, and the University of the Western Cape), the Computational Biology division at UCT and the Data Intensive Research Initiative of South Africa (DIRISA).



\bibliographystyle{mnras}
\bibliography{example} % if your bibtex file is called example.bib

\begin{thebibliography}{}
\makeatletter
\relax
\def\mn@urlcharsother{\let\do\@makeother \do\$\do\&\do\#\do\^\do\_\do\%\do\~}
\def\mn@doi{\begingroup\mn@urlcharsother \@ifnextchar [ {\mn@doi@}
  {\mn@doi@[]}}
\def\mn@doi@[#1]#2{\def\@tempa{#1}\ifx\@tempa\@empty \href
  {http://dx.doi.org/#2} {doi:#2}\else \href {http://dx.doi.org/#2} {#1}\fi
  \endgroup}
\def\mn@eprint#1#2{\mn@eprint@#1:#2::\@nil}
\def\mn@eprint@arXiv#1{\href {http://arxiv.org/abs/#1} {{\tt arXiv:#1}}}
\def\mn@eprint@dblp#1{\href {http://dblp.uni-trier.de/rec/bibtex/#1.xml}
  {dblp:#1}}
\def\mn@eprint@#1:#2:#3:#4\@nil{\def\@tempa {#1}\def\@tempb {#2}\def\@tempc
  {#3}\ifx \@tempc \@empty \let \@tempc \@tempb \let \@tempb \@tempa \fi \ifx
  \@tempb \@empty \def\@tempb {arXiv}\fi \@ifundefined
  {mn@eprint@\@tempb}{\@tempb:\@tempc}{\expandafter \expandafter \csname
  mn@eprint@\@tempb\endcsname \expandafter{\@tempc}}}

\bibitem[\protect\citeauthoryear{{Abraham}}{{Abraham}}{2000}]{2000A&A...355..915A}
{Abraham} Z.,  2000, \aap, \href
  {https://ui.adsabs.harvard.edu/abs/2000A&A...355..915A} {355, 915}

\bibitem[\protect\citeauthoryear{{Agazie} et~al.,}{{Agazie}
  et~al.}{2024}]{2024ApJ...963..144A}
{Agazie} G.,  et~al., 2024, \mn@doi [\apj] {10.3847/1538-4357/ad1f61}, \href
  {https://ui.adsabs.harvard.edu/abs/2024ApJ...963..144A} {963, 144}

\bibitem[\protect\citeauthoryear{{Bansal}, {Taylor}, {Peck}, {Zavala}  \&
  {Romani}}{{Bansal} et~al.}{2017}]{2017ApJ...843...14B}
{Bansal} K.,  {Taylor} G.~B.,  {Peck} A.~B.,  {Zavala} R.~T.,   {Romani} R.~W.,
   2017, \mn@doi [\apj] {10.3847/1538-4357/aa74e1}, \href
  {https://ui.adsabs.harvard.edu/abs/2017ApJ...843...14B} {843, 14}

\bibitem[\protect\citeauthoryear{{B{\'e}csy}, {Cornish}, {Petrov}, {Siemens},
  {Taylor}, {Vigeland}  \& {Witt}}{{B{\'e}csy}
  et~al.}{2025}]{2025CQGra..42q5016B}
{B{\'e}csy} B.,  {Cornish} N.~J.,  {Petrov} P.,  {Siemens} X.,  {Taylor} S.~R.,
   {Vigeland} S.~J.,   {Witt} C.~A.,  2025, \mn@doi [Classical and Quantum
  Gravity] {10.1088/1361-6382/adfd36}, \href
  {https://ui.adsabs.harvard.edu/abs/2025CQGra..42q5016B} {42, 175016}

\bibitem[\protect\citeauthoryear{{Begelman}, {Blandford}  \& {Rees}}{{Begelman}
  et~al.}{1980}]{1980Natur.287..307B}
{Begelman} M.~C.,  {Blandford} R.~D.,   {Rees} M.~J.,  1980, \mn@doi [\nat]
  {10.1038/287307a0}, \href
  {https://ui.adsabs.harvard.edu/abs/1980Natur.287..307B} {287, 307}

\bibitem[\protect\citeauthoryear{{Bhatta}}{{Bhatta}}{2018}]{2018Galax...6..136B}
{Bhatta} G.,  2018, \mn@doi [Galaxies] {10.3390/galaxies6040136}, \href
  {https://ui.adsabs.harvard.edu/abs/2018Galax...6..136B} {6, 136}

\bibitem[\protect\citeauthoryear{{Bian}, {Ge}, {Shu}, {Wang}, {Yang}  \&
  {Zong}}{{Bian} et~al.}{2024}]{2024PhRvD.109j1301B}
{Bian} L.,  {Ge} S.,  {Shu} J.,  {Wang} B.,  {Yang} X.-Y.,   {Zong} J.,  2024,
  \mn@doi [\prd] {10.1103/PhysRevD.109.L101301}, \href
  {https://ui.adsabs.harvard.edu/abs/2024PhRvD.109j1301B} {109, L101301}

\bibitem[\protect\citeauthoryear{{Bianchi}, {Chiaberge}, {Piconcelli},
  {Guainazzi}  \& {Matt}}{{Bianchi} et~al.}{2008}]{2008MNRAS.386..105B}
{Bianchi} S.,  {Chiaberge} M.,  {Piconcelli} E.,  {Guainazzi} M.,   {Matt} G.,
  2008, \mn@doi [\mnras] {10.1111/j.1365-2966.2008.13078.x}, \href
  {https://ui.adsabs.harvard.edu/abs/2008MNRAS.386..105B} {386, 105}

\bibitem[\protect\citeauthoryear{{Britzen} et~al.,}{{Britzen}
  et~al.}{2018}]{2018MNRAS.478.3199B}
{Britzen} S.,  et~al., 2018, \mn@doi [\mnras] {10.1093/mnras/sty1026}, \href
  {https://ui.adsabs.harvard.edu/abs/2018MNRAS.478.3199B} {478, 3199}

\bibitem[\protect\citeauthoryear{{Britzen}, {Zaja{\v{c}}ek}, {Gopal-Krishna},
  {Fendt}, {Kun}, {Jaron}, {Sillanp{\"a}{\"a}}  \& {Eckart}}{{Britzen}
  et~al.}{2023}]{2023ApJ...951..106B}
{Britzen} S.,  {Zaja{\v{c}}ek} M.,  {Gopal-Krishna} {Fendt} C.,  {Kun} E.,
  {Jaron} F.,  {Sillanp{\"a}{\"a}} A.,   {Eckart} A.,  2023, \mn@doi [\apj]
  {10.3847/1538-4357/accbbc}, \href
  {https://ui.adsabs.harvard.edu/abs/2023ApJ...951..106B} {951, 106}

\bibitem[\protect\citeauthoryear{{Burke-Spolaor} et~al.,}{{Burke-Spolaor}
  et~al.}{2019}]{2019A&ARv..27....5B}
{Burke-Spolaor} S.,  et~al., 2019, \mn@doi [\aapr] {10.1007/s00159-019-0115-7},
  \href {https://ui.adsabs.harvard.edu/abs/2019A&ARv..27....5B} {27, 5}

\bibitem[\protect\citeauthoryear{{Caproni} \& {Abraham}}{{Caproni} \&
  {Abraham}}{2004}]{2004ApJ...602..625C}
{Caproni} A.,  {Abraham} Z.,  2004, \mn@doi [\apj] {10.1086/381195}, \href
  {https://ui.adsabs.harvard.edu/abs/2004ApJ...602..625C} {602, 625}

\bibitem[\protect\citeauthoryear{{Caproni}, {Abraham}  \& {Monteiro}}{{Caproni}
  et~al.}{2013}]{2013MNRAS.428..280C}
{Caproni} A.,  {Abraham} Z.,   {Monteiro} H.,  2013, \mn@doi [\mnras]
  {10.1093/mnras/sts014}, \href
  {https://ui.adsabs.harvard.edu/abs/2013MNRAS.428..280C} {428, 280}

\bibitem[\protect\citeauthoryear{{Caproni}, {Abraham}, {Motter}  \&
  {Monteiro}}{{Caproni} et~al.}{2017}]{2017ApJ...851L..39C}
{Caproni} A.,  {Abraham} Z.,  {Motter} J.~C.,   {Monteiro} H.,  2017, \mn@doi
  [\apjl] {10.3847/2041-8213/aa9fea}, \href
  {https://ui.adsabs.harvard.edu/abs/2017ApJ...851L..39C} {851, L39}

\bibitem[\protect\citeauthoryear{{Casey-Clyde}, {Mingarelli}, {Greene},
  {Pardo}, {Na{\~n}ez}  \& {Goulding}}{{Casey-Clyde}
  et~al.}{2022}]{2022ApJ...924...93C}
{Casey-Clyde} J.~A.,  {Mingarelli} C. M.~F.,  {Greene} J.~E.,  {Pardo} K.,
  {Na{\~n}ez} M.,   {Goulding} A.~D.,  2022, \mn@doi [\apj]
  {10.3847/1538-4357/ac32de}, \href
  {https://ui.adsabs.harvard.edu/abs/2022ApJ...924...93C} {924, 93}

\bibitem[\protect\citeauthoryear{{Charisi}, {Bartos}, {Haiman}, {Price-Whelan},
  {Graham}, {Bellm}, {Laher}  \& {M{\'a}rka}}{{Charisi}
  et~al.}{2016}]{2016MNRAS.463.2145C}
{Charisi} M.,  {Bartos} I.,  {Haiman} Z.,  {Price-Whelan} A.~M.,  {Graham}
  M.~J.,  {Bellm} E.~C.,  {Laher} R.~R.,   {M{\'a}rka} S.,  2016, \mn@doi
  [\mnras] {10.1093/mnras/stw1838}, \href
  {https://ui.adsabs.harvard.edu/abs/2016MNRAS.463.2145C} {463, 2145}

\bibitem[\protect\citeauthoryear{{Chen} et~al.,}{{Chen}
  et~al.}{2020}]{2020MNRAS.499.2245C}
{Chen} Y.-C.,  et~al., 2020, \mn@doi [\mnras] {10.1093/mnras/staa2957}, \href
  {https://ui.adsabs.harvard.edu/abs/2020MNRAS.499.2245C} {499, 2245}

\bibitem[\protect\citeauthoryear{{Cho} et~al.,}{{Cho}
  et~al.}{2024}]{2024A&A...683A.248C}
{Cho} I.,  et~al., 2024, \mn@doi [\aap] {10.1051/0004-6361/202347157}, \href
  {https://ui.adsabs.harvard.edu/abs/2024A&A...683A.248C} {683, A248}

\bibitem[\protect\citeauthoryear{{Comerford}, {Pooley}, {Barrows}, {Greene},
  {Zakamska}, {Madejski}  \& {Cooper}}{{Comerford}
  et~al.}{2015}]{2015ApJ...806..219C}
{Comerford} J.~M.,  {Pooley} D.,  {Barrows} R.~S.,  {Greene} J.~E.,  {Zakamska}
  N.~L.,  {Madejski} G.~M.,   {Cooper} M.~C.,  2015, \mn@doi [\apj]
  {10.1088/0004-637X/806/2/219}, \href
  {https://ui.adsabs.harvard.edu/abs/2015ApJ...806..219C} {806, 219}

\bibitem[\protect\citeauthoryear{{Conway} \& {Murphy}}{{Conway} \&
  {Murphy}}{1993}]{1993ApJ...411...89C}
{Conway} J.~E.,  {Murphy} D.~W.,  1993, \mn@doi [\apj] {10.1086/172809}, \href
  {https://ui.adsabs.harvard.edu/abs/1993ApJ...411...89C} {411, 89}

\bibitem[\protect\citeauthoryear{{D'Orazio} \& {Charisi}}{{D'Orazio} \&
  {Charisi}}{2023}]{2023arXiv231016896D}
{D'Orazio} D.~J.,  {Charisi} M.,  2023, \mn@doi [arXiv e-prints]
  {10.48550/arXiv.2310.16896}, \href
  {https://ui.adsabs.harvard.edu/abs/2023arXiv231016896D} {p. arXiv:2310.16896}

\bibitem[\protect\citeauthoryear{{D'Orazio}, {Haiman}  \&
  {MacFadyen}}{{D'Orazio} et~al.}{2013}]{2013MNRAS.436.2997D}
{D'Orazio} D.~J.,  {Haiman} Z.,   {MacFadyen} A.,  2013, \mn@doi [\mnras]
  {10.1093/mnras/stt1787}, \href
  {https://ui.adsabs.harvard.edu/abs/2013MNRAS.436.2997D} {436, 2997}

\bibitem[\protect\citeauthoryear{{D'Orazio}, {Haiman}  \&
  {Schiminovich}}{{D'Orazio} et~al.}{2015}]{2015Natur.525..351D}
{D'Orazio} D.~J.,  {Haiman} Z.,   {Schiminovich} D.,  2015, \mn@doi [\nat]
  {10.1038/nature15262}, \href
  {https://ui.adsabs.harvard.edu/abs/2015Natur.525..351D} {525, 351}

\bibitem[\protect\citeauthoryear{{De Rosa} et~al.,}{{De Rosa}
  et~al.}{2019}]{2019NewAR..8601525D}
{De Rosa} A.,  et~al., 2019, \mn@doi [\nar] {10.1016/j.newar.2020.101525},
  \href {https://ui.adsabs.harvard.edu/abs/2019NewAR..8601525D} {86, 101525}

\bibitem[\protect\citeauthoryear{{Deng}, {Xiang}, {Chen}, {Jing}, {Zhu}  \&
  {Wu}}{{Deng} et~al.}{2026}]{2026ApJS..283...70D}
{Deng} Z.,  {Xiang} C.,  {Chen} Q.,  {Jing} L.,  {Zhu} X.,   {Wu} J.,  2026,
  \mn@doi [\apjs] {10.3847/1538-4365/ae497a}, \href
  {https://ui.adsabs.harvard.edu/abs/2026ApJS..283...70D} {283, 70}

\bibitem[\protect\citeauthoryear{{Dou} et~al.,}{{Dou}
  et~al.}{2022}]{2022A&A...665L...3D}
{Dou} L.,  et~al., 2022, \mn@doi [\aap] {10.1051/0004-6361/202244450}, \href
  {https://ui.adsabs.harvard.edu/abs/2022A&A...665L...3D} {665, L3}

\bibitem[\protect\citeauthoryear{{Drake} et~al.,}{{Drake}
  et~al.}{2009}]{2009ApJ...696..870D}
{Drake} A.~J.,  et~al., 2009, \mn@doi [\apj] {10.1088/0004-637X/696/1/870},
  \href {https://ui.adsabs.harvard.edu/abs/2009ApJ...696..870D} {696, 870}

\bibitem[\protect\citeauthoryear{{Farris}, {Duffell}, {MacFadyen}  \&
  {Haiman}}{{Farris} et~al.}{2014}]{2014ApJ...783..134F}
{Farris} B.~D.,  {Duffell} P.,  {MacFadyen} A.~I.,   {Haiman} Z.,  2014,
  \mn@doi [\apj] {10.1088/0004-637X/783/2/134}, \href
  {https://ui.adsabs.harvard.edu/abs/2014ApJ...783..134F} {783, 134}

\bibitem[\protect\citeauthoryear{{Fu} et~al.,}{{Fu}
  et~al.}{2011}]{2011ApJ...740L..44F}
{Fu} H.,  et~al., 2011, \mn@doi [\apjl] {10.1088/2041-8205/740/2/L44}, \href
  {https://ui.adsabs.harvard.edu/abs/2011ApJ...740L..44F} {740, L44}

\bibitem[\protect\citeauthoryear{{Gold}, {Paschalidis}, {Etienne}, {Shapiro}
  \& {Pfeiffer}}{{Gold} et~al.}{2014}]{2014PhRvD..89f4060G}
{Gold} R.,  {Paschalidis} V.,  {Etienne} Z.~B.,  {Shapiro} S.~L.,   {Pfeiffer}
  H.~P.,  2014, \mn@doi [\prd] {10.1103/PhysRevD.89.064060}, \href
  {https://ui.adsabs.harvard.edu/abs/2014PhRvD..89f4060G} {89, 064060}

\bibitem[\protect\citeauthoryear{{Graham} et~al.,}{{Graham}
  et~al.}{2015a}]{2015MNRAS.453.1562G}
{Graham} M.~J.,  et~al., 2015a, \mn@doi [\mnras] {10.1093/mnras/stv1726}, \href
  {https://ui.adsabs.harvard.edu/abs/2015MNRAS.453.1562G} {453, 1562}

\bibitem[\protect\citeauthoryear{{Graham} et~al.,}{{Graham}
  et~al.}{2015b}]{graham2015}
{Graham} M.~J.,  et~al., 2015b, \mn@doi [\nat] {10.1038/nature14143}, \href
  {https://ui.adsabs.harvard.edu/abs/2015Natur.518...74G} {518, 74}

\bibitem[\protect\citeauthoryear{{Guyon}, {Sanders}  \& {Stockton}}{{Guyon}
  et~al.}{2006}]{2006ApJS..166...89G}
{Guyon} O.,  {Sanders} D.~B.,   {Stockton} A.,  2006, \mn@doi [\apjs]
  {10.1086/505030}, \href
  {https://ui.adsabs.harvard.edu/abs/2006ApJS..166...89G} {166, 89}

\bibitem[\protect\citeauthoryear{{Hong}, {Im}, {Kim}  \& {Ho}}{{Hong}
  et~al.}{2015}]{2015ApJ...804...34H}
{Hong} J.,  {Im} M.,  {Kim} M.,   {Ho} L.~C.,  2015, \mn@doi [\apj]
  {10.1088/0004-637X/804/1/34}, \href
  {https://ui.adsabs.harvard.edu/abs/2015ApJ...804...34H} {804, 34}

\bibitem[\protect\citeauthoryear{{Johnson} et~al.,}{{Johnson}
  et~al.}{2024}]{2024PhRvD.109j3012J}
{Johnson} A.~D.,  et~al., 2024, \mn@doi [\prd] {10.1103/PhysRevD.109.103012},
  \href {https://ui.adsabs.harvard.edu/abs/2024PhRvD.109j3012J} {109, 103012}

\bibitem[\protect\citeauthoryear{{Jun}, {Stern}, {Graham}, {Djorgovski},
  {Mainzer}, {Cutri}, {Drake}  \& {Mahabal}}{{Jun}
  et~al.}{2015}]{2015ApJ...814L..12J}
{Jun} H.~D.,  {Stern} D.,  {Graham} M.~J.,  {Djorgovski} S.~G.,  {Mainzer} A.,
  {Cutri} R.~M.,  {Drake} A.~J.,   {Mahabal} A.~A.,  2015, \mn@doi [\apjl]
  {10.1088/2041-8205/814/1/L12}, \href
  {https://ui.adsabs.harvard.edu/abs/2015ApJ...814L..12J} {814, L12}

\bibitem[\protect\citeauthoryear{{Kauffmann} \& {Haehnelt}}{{Kauffmann} \&
  {Haehnelt}}{2000}]{2000MNRAS.311..576K}
{Kauffmann} G.,  {Haehnelt} M.,  2000, \mn@doi [\mnras]
  {10.1046/j.1365-8711.2000.03077.x}, \href
  {https://ui.adsabs.harvard.edu/abs/2000MNRAS.311..576K} {311, 576}

\bibitem[\protect\citeauthoryear{{Kharb}, {Lal}  \& {Merritt}}{{Kharb}
  et~al.}{2017}]{2017NatAs...1..727K}
{Kharb} P.,  {Lal} D.~V.,   {Merritt} D.,  2017, \mn@doi [Nature Astronomy]
  {10.1038/s41550-017-0256-4}, \href
  {https://ui.adsabs.harvard.edu/abs/2017NatAs...1..727K} {1, 727}

\bibitem[\protect\citeauthoryear{{Kochanek} et~al.,}{{Kochanek}
  et~al.}{2017}]{2017PASP..129j4502K}
{Kochanek} C.~S.,  et~al., 2017, \mn@doi [\pasp] {10.1088/1538-3873/aa80d9},
  \href {https://ui.adsabs.harvard.edu/abs/2017PASP..129j4502K} {129, 104502}

\bibitem[\protect\citeauthoryear{{Komossa}, {Burwitz}, {Hasinger}, {Predehl},
  {Kaastra}  \& {Ikebe}}{{Komossa} et~al.}{2003}]{2003ApJ...582L..15K}
{Komossa} S.,  {Burwitz} V.,  {Hasinger} G.,  {Predehl} P.,  {Kaastra} J.~S.,
  {Ikebe} Y.,  2003, \mn@doi [\apjl] {10.1086/346145}, \href
  {https://ui.adsabs.harvard.edu/abs/2003ApJ...582L..15K} {582, L15}

\bibitem[\protect\citeauthoryear{{Kormendy} \& {Ho}}{{Kormendy} \&
  {Ho}}{2013}]{2013ARA&A..51..511K}
{Kormendy} J.,  {Ho} L.~C.,  2013, \mn@doi [\araa]
  {10.1146/annurev-astro-082708-101811}, \href
  {https://ui.adsabs.harvard.edu/abs/2013ARA&A..51..511K} {51, 511}

\bibitem[\protect\citeauthoryear{{Koss} et~al.,}{{Koss}
  et~al.}{2018}]{2018Natur.563..214K}
{Koss} M.~J.,  et~al., 2018, \mn@doi [\nat] {10.1038/s41586-018-0652-7}, \href
  {https://ui.adsabs.harvard.edu/abs/2018Natur.563..214K} {563, 214}

\bibitem[\protect\citeauthoryear{{Kun}, {Gab{\'a}nyi}, {Karouzos}, {Britzen}
  \& {Gergely}}{{Kun} et~al.}{2014}]{2014MNRAS.445.1370K}
{Kun} E.,  {Gab{\'a}nyi} K.~{\'E}.,  {Karouzos} M.,  {Britzen} S.,   {Gergely}
  L.~{\'A}.,  2014, \mn@doi [\mnras] {10.1093/mnras/stu1813}, \href
  {https://ui.adsabs.harvard.edu/abs/2014MNRAS.445.1370K} {445, 1370}

\bibitem[\protect\citeauthoryear{{Kun}, {Frey}, {Gab{\'a}nyi}, {Britzen},
  {Cseh}  \& {Gergely}}{{Kun} et~al.}{2015}]{2015MNRAS.454.1290K}
{Kun} E.,  {Frey} S.,  {Gab{\'a}nyi} K.~{\'E}.,  {Britzen} S.,  {Cseh} D.,
  {Gergely} L.~{\'A}.,  2015, \mn@doi [\mnras] {10.1093/mnras/stv2049}, \href
  {https://ui.adsabs.harvard.edu/abs/2015MNRAS.454.1290K} {454, 1290}

\bibitem[\protect\citeauthoryear{{Leroy}, {Bobin}  \& {Moutarde}}{{Leroy}
  et~al.}{2024}]{2024A&A...689A.107L}
{Leroy} E.,  {Bobin} J.,   {Moutarde} H.,  2024, \mn@doi [\aap]
  {10.1051/0004-6361/202449987}, \href
  {https://ui.adsabs.harvard.edu/abs/2024A&A...689A.107L} {689, A107}

\bibitem[\protect\citeauthoryear{{Li}, {Zhuang}, {Shen}, {Volonteri}, {Chen}
  \& {Matteo}}{{Li} et~al.}{2025}]{2025ApJ...986..101L}
{Li} J.,  {Zhuang} M.-Y.,  {Shen} Y.,  {Volonteri} M.,  {Chen} N.,   {Matteo}
  T.~D.,  2025, \mn@doi [\apj] {10.3847/1538-4357/adbae2}, \href
  {https://ui.adsabs.harvard.edu/abs/2025ApJ...986..101L} {986, 101}

\bibitem[\protect\citeauthoryear{{Liao} et~al.,}{{Liao}
  et~al.}{2021}]{2021MNRAS.500.4025L}
{Liao} W.-T.,  et~al., 2021, \mn@doi [\mnras] {10.1093/mnras/staa3055}, \href
  {https://ui.adsabs.harvard.edu/abs/2021MNRAS.500.4025L} {500, 4025}

\bibitem[\protect\citeauthoryear{{Liu} et~al.,}{{Liu}
  et~al.}{2015}]{2015ApJ...803L..16L}
{Liu} T.,  et~al., 2015, \mn@doi [\apjl] {10.1088/2041-8205/803/2/L16}, \href
  {https://ui.adsabs.harvard.edu/abs/2015ApJ...803L..16L} {803, L16}

\bibitem[\protect\citeauthoryear{{Liu} et~al.,}{{Liu}
  et~al.}{2016}]{2016ApJ...833....6L}
{Liu} T.,  et~al., 2016, \mn@doi [\apj] {10.3847/0004-637X/833/1/6}, \href
  {https://ui.adsabs.harvard.edu/abs/2016ApJ...833....6L} {833, 6}

\bibitem[\protect\citeauthoryear{{Liu}, {Gezari}  \& {Miller}}{{Liu}
  et~al.}{2018}]{2018ApJ...859L..12L}
{Liu} T.,  {Gezari} S.,   {Miller} M.~C.,  2018, \mn@doi [\apjl]
  {10.3847/2041-8213/aac2ed}, \href
  {https://ui.adsabs.harvard.edu/abs/2018ApJ...859L..12L} {859, L12}

\bibitem[\protect\citeauthoryear{{Liu} et~al.,}{{Liu}
  et~al.}{2024}]{2024ApJ...964..167L}
{Liu} T.,  et~al., 2024, \mn@doi [\apj] {10.3847/1538-4357/ad23e2}, \href
  {https://ui.adsabs.harvard.edu/abs/2024ApJ...964..167L} {964, 167}

\bibitem[\protect\citeauthoryear{{Lomb}}{{Lomb}}{1976}]{1976Ap&SS..39..447L}
{Lomb} N.~R.,  1976, \mn@doi [\apss] {10.1007/BF00648343}, \href
  {https://ui.adsabs.harvard.edu/abs/1976Ap&SS..39..447L} {39, 447}

\bibitem[\protect\citeauthoryear{{Lu}, {Chiang}  \& {Li}}{{Lu}
  et~al.}{2024}]{2024PhRvD.109j1304L}
{Lu} B.-Q.,  {Chiang} C.-W.,   {Li} T.,  2024, \mn@doi [\prd]
  {10.1103/PhysRevD.109.L101304}, \href
  {https://ui.adsabs.harvard.edu/abs/2024PhRvD.109j1304L} {109, L101304}

\bibitem[\protect\citeauthoryear{{Manchester} \& {IPTA}}{{Manchester} \&
  {IPTA}}{2013}]{2013CQGra..30v4010M}
{Manchester} R.~N.,  {IPTA} 2013, \mn@doi [Classical and Quantum Gravity]
  {10.1088/0264-9381/30/22/224010}, \href
  {https://ui.adsabs.harvard.edu/abs/2013CQGra..30v4010M} {30, 224010}

\bibitem[\protect\citeauthoryear{{Martin} et~al.,}{{Martin}
  et~al.}{2005}]{2005ApJ...619L...1M}
{Martin} D.~C.,  et~al., 2005, \mn@doi [\apjl] {10.1086/426387}, \href
  {https://ui.adsabs.harvard.edu/abs/2005ApJ...619L...1M} {619, L1}

\bibitem[\protect\citeauthoryear{{Marziani}, {Sulentic}, {Dultzin-Hacyan},
  {Calvani}  \& {Moles}}{{Marziani} et~al.}{1996}]{1996ApJS..104...37M}
{Marziani} P.,  {Sulentic} J.~W.,  {Dultzin-Hacyan} D.,  {Calvani} M.,
  {Moles} M.,  1996, \mn@doi [\apjs] {10.1086/192291}, \href
  {https://ui.adsabs.harvard.edu/abs/1996ApJS..104...37M} {104, 37}

\bibitem[\protect\citeauthoryear{{Max-Moerbeck}, {Richards}, {Hovatta},
  {Pavlidou}, {Pearson}  \& {Readhead}}{{Max-Moerbeck}
  et~al.}{2014}]{2014MNRAS.445..437M}
{Max-Moerbeck} W.,  {Richards} J.~L.,  {Hovatta} T.,  {Pavlidou} V.,  {Pearson}
  T.~J.,   {Readhead} A.~C.~S.,  2014, \mn@doi [\mnras]
  {10.1093/mnras/stu1707}, \href
  {https://ui.adsabs.harvard.edu/abs/2014MNRAS.445..437M} {445, 437}

\bibitem[\protect\citeauthoryear{{Mohan} \& {Mangalam}}{{Mohan} \&
  {Mangalam}}{2015}]{2015ApJ...805...91M}
{Mohan} P.,  {Mangalam} A.,  2015, \mn@doi [\apj] {10.1088/0004-637X/805/2/91},
  \href {https://ui.adsabs.harvard.edu/abs/2015ApJ...805...91M} {805, 91}

\bibitem[\protect\citeauthoryear{{Mohan}, {An}, {Frey}, {Mangalam},
  {Gab{\'a}nyi}  \& {Kun}}{{Mohan} et~al.}{2016}]{2016MNRAS.463.1812M}
{Mohan} P.,  {An} T.,  {Frey} S.,  {Mangalam} A.,  {Gab{\'a}nyi} K.~{\'E}.,
  {Kun} E.,  2016, \mn@doi [\mnras] {10.1093/mnras/stw2154}, \href
  {https://ui.adsabs.harvard.edu/abs/2016MNRAS.463.1812M} {463, 1812}

\bibitem[\protect\citeauthoryear{{Planck Collaboration}}{{Planck
  Collaboration}}{2018}]{Planck2018}
{Planck Collaboration} 2018, \mn@doi
  [\href{https://www.aanda.org/articles/aa/abs/2020/09/aa36904-20/aa36904-20.html}{Astron.
  Astrophys.}] {10.1051/0004-6361/201833910}, 641, A6

\bibitem[\protect\citeauthoryear{{Prince}, {Banerjee}, {Sharma}, {Kumar das},
  {Gupta}  \& {Bose}}{{Prince} et~al.}{2023}]{2023A&A...678A.100P}
{Prince} R.,  {Banerjee} A.,  {Sharma} A.,  {Kumar das} A.,  {Gupta} A.~C.,
  {Bose} D.,  2023, \mn@doi [\aap] {10.1051/0004-6361/202346400}, \href
  {https://ui.adsabs.harvard.edu/abs/2023A&A...678A.100P} {678, A100}

\bibitem[\protect\citeauthoryear{{Qian}, {Britzen}, {Witzel}, {Krichbaum}  \&
  {Kun}}{{Qian} et~al.}{2018}]{2018A&A...615A.123Q}
{Qian} S.~J.,  {Britzen} S.,  {Witzel} A.,  {Krichbaum} T.~P.,   {Kun} E.,
  2018, \mn@doi [\aap] {10.1051/0004-6361/201732039}, \href
  {https://ui.adsabs.harvard.edu/abs/2018A&A...615A.123Q} {615, A123}

\bibitem[\protect\citeauthoryear{{Rigamonti} et~al.,}{{Rigamonti}
  et~al.}{2025}]{2025A&A...693A.117R}
{Rigamonti} F.,  et~al., 2025, \mn@doi [\aap] {10.1051/0004-6361/202452830},
  \href {https://ui.adsabs.harvard.edu/abs/2025A&A...693A.117R} {693, A117}

\bibitem[\protect\citeauthoryear{{Rodriguez}, {Taylor}, {Zavala}, {Peck},
  {Pollack}  \& {Romani}}{{Rodriguez} et~al.}{2006}]{2006ApJ...646...49R}
{Rodriguez} C.,  {Taylor} G.~B.,  {Zavala} R.~T.,  {Peck} A.~B.,  {Pollack}
  L.~K.,   {Romani} R.~W.,  2006, \mn@doi [\apj] {10.1086/504825}, \href
  {https://ui.adsabs.harvard.edu/abs/2006ApJ...646...49R} {646, 49}

\bibitem[\protect\citeauthoryear{{Romero}, {Chajet}, {Abraham}  \&
  {Fan}}{{Romero} et~al.}{2000}]{2000A&A...360...57R}
{Romero} G.~E.,  {Chajet} L.,  {Abraham} Z.,   {Fan} J.~H.,  2000, \aap, \href
  {https://ui.adsabs.harvard.edu/abs/2000A&A...360...57R} {360, 57}

\bibitem[\protect\citeauthoryear{{Romero}, {Vila}  \& {P{\'e}rez}}{{Romero}
  et~al.}{2016}]{2016A&A...588A.125R}
{Romero} G.~E.,  {Vila} G.~S.,   {P{\'e}rez} D.,  2016, \mn@doi [\aap]
  {10.1051/0004-6361/201527479}, \href
  {https://ui.adsabs.harvard.edu/abs/2016A&A...588A.125R} {588, A125}

\bibitem[\protect\citeauthoryear{{Sault}, {Teuben}  \& {Wright}}{{Sault}
  et~al.}{1995}]{miriad}
{Sault} R.~J.,  {Teuben} P.~J.,   {Wright} M.~C.~H.,  1995, in {Shaw} R.~A.,
  {Payne} H.~E.,   {Hayes} J.~J.~E.,  eds,  Astronomical Society of the Pacific
  Conference Series Vol. 77, Astronomical Data Analysis Software and Systems
  IV. p.~433 (\mn@eprint {arXiv} {astro-ph/0612759}),
  \mn@doi{10.48550/arXiv.astro-ph/0612759}

\bibitem[\protect\citeauthoryear{{Scargle}}{{Scargle}}{1982}]{1982ApJ...263..835S}
{Scargle} J.~D.,  1982, \mn@doi [\apj] {10.1086/160554}, \href
  {https://ui.adsabs.harvard.edu/abs/1982ApJ...263..835S} {263, 835}

\bibitem[\protect\citeauthoryear{{Sesana}, {Vecchio}  \& {Volonteri}}{{Sesana}
  et~al.}{2009}]{2009MNRAS.394.2255S}
{Sesana} A.,  {Vecchio} A.,   {Volonteri} M.,  2009, \mn@doi [\mnras]
  {10.1111/j.1365-2966.2009.14499.x}, \href
  {https://ui.adsabs.harvard.edu/abs/2009MNRAS.394.2255S} {394, 2255}

\bibitem[\protect\citeauthoryear{{Sesar}, {Stuart}, {Ivezi{\'c}}, {Morgan},
  {Becker}  \& {Wo{\'z}niak}}{{Sesar} et~al.}{2011}]{2011AJ....142..190S}
{Sesar} B.,  {Stuart} J.~S.,  {Ivezi{\'c}} {\v{Z}}.,  {Morgan} D.~P.,  {Becker}
  A.~C.,   {Wo{\'z}niak} P.,  2011, \mn@doi [\aj]
  {10.1088/0004-6256/142/6/190}, \href
  {https://ui.adsabs.harvard.edu/abs/2011AJ....142..190S} {142, 190}

\bibitem[\protect\citeauthoryear{{Shappee} et~al.,}{{Shappee}
  et~al.}{2014}]{2014ApJ...788...48S}
{Shappee} B.~J.,  et~al., 2014, \mn@doi [\apj] {10.1088/0004-637X/788/1/48},
  \href {https://ui.adsabs.harvard.edu/abs/2014ApJ...788...48S} {788, 48}

\bibitem[\protect\citeauthoryear{{Stirling} et~al.,}{{Stirling}
  et~al.}{2003}]{2003MNRAS.341..405S}
{Stirling} A.~M.,  et~al., 2003, \mn@doi [\mnras]
  {10.1046/j.1365-8711.2003.06448.x}, \href
  {https://ui.adsabs.harvard.edu/abs/2003MNRAS.341..405S} {341, 405}

\bibitem[\protect\citeauthoryear{{Sudou}, {Iguchi}, {Murata}  \&
  {Taniguchi}}{{Sudou} et~al.}{2003}]{2003Sci...300.1263S}
{Sudou} H.,  {Iguchi} S.,  {Murata} Y.,   {Taniguchi} Y.,  2003, \mn@doi
  [Science] {10.1126/science.1082817}, \href
  {https://ui.adsabs.harvard.edu/abs/2003Sci...300.1263S} {300, 1263}

\bibitem[\protect\citeauthoryear{{Tateyama}, {Kingham}, {Kaufmann}, {Piner},
  {de Lucena}  \& {Botti}}{{Tateyama} et~al.}{1998}]{1998ApJ...500..810T}
{Tateyama} C.~E.,  {Kingham} K.~A.,  {Kaufmann} P.,  {Piner} B.~G.,  {de
  Lucena} A.~M.~P.,   {Botti} L.~C.~L.,  1998, \mn@doi [\apj] {10.1086/305783},
  \href {https://ui.adsabs.harvard.edu/abs/1998ApJ...500..810T} {500, 810}

\bibitem[\protect\citeauthoryear{{Urry} \& {Padovani}}{{Urry} \&
  {Padovani}}{1995}]{1995PASP..107..803U}
{Urry} C.~M.,  {Padovani} P.,  1995, \mn@doi [\pasp] {10.1086/133630}, \href
  {https://ui.adsabs.harvard.edu/abs/1995PASP..107..803U} {107, 803}

\bibitem[\protect\citeauthoryear{{VanderPlas}}{{VanderPlas}}{2018}]{2018ApJS..236...16V}
{VanderPlas} J.~T.,  2018, \mn@doi [\apjs] {10.3847/1538-4365/aab766}, \href
  {https://ui.adsabs.harvard.edu/abs/2018ApJS..236...16V} {236, 16}

\bibitem[\protect\citeauthoryear{{Vaughan}}{{Vaughan}}{2005}]{2005A&A...431..391V}
{Vaughan} S.,  2005, \mn@doi [\aap] {10.1051/0004-6361:20041453}, \href
  {https://ui.adsabs.harvard.edu/abs/2005A&A...431..391V} {431, 391}

\bibitem[\protect\citeauthoryear{{Vaughan} \& {Uttley}}{{Vaughan} \&
  {Uttley}}{2005}]{2005MNRAS.362..235V}
{Vaughan} S.,  {Uttley} P.,  2005, \mn@doi [\mnras]
  {10.1111/j.1365-2966.2005.09296.x}, \href
  {https://ui.adsabs.harvard.edu/abs/2005MNRAS.362..235V} {362, 235}

\bibitem[\protect\citeauthoryear{{Woo}, {Cho}, {Husemann}, {Komossa}, {Park}
  \& {Bennert}}{{Woo} et~al.}{2014}]{2014MNRAS.437...32W}
{Woo} J.-H.,  {Cho} H.,  {Husemann} B.,  {Komossa} S.,  {Park} D.,   {Bennert}
  V.~N.,  2014, \mn@doi [\mnras] {10.1093/mnras/stt1846}, \href
  {https://ui.adsabs.harvard.edu/abs/2014MNRAS.437...32W} {437, 32}

\bibitem[\protect\citeauthoryear{{Zheng}, {Butler}, {Shen}, {Jiang}, {Wang},
  {Chen}  \& {Cuadra}}{{Zheng} et~al.}{2016}]{2016ApJ...827...56Z}
{Zheng} Z.-Y.,  {Butler} N.~R.,  {Shen} Y.,  {Jiang} L.,  {Wang} J.-X.,  {Chen}
  X.,   {Cuadra} J.,  2016, \mn@doi [\apj] {10.3847/0004-637X/827/1/56}, \href
  {https://ui.adsabs.harvard.edu/abs/2016ApJ...827...56Z} {827, 56}

\bibitem[\protect\citeauthoryear{{Zheng}, {Zhang}, {Yuan}, {Severgnini}  \&
  {Vignali}}{{Zheng} et~al.}{2024}]{2024MNRAS.531L..76Z}
{Zheng} Q.,  {Zhang} X.,  {Yuan} Q.,  {Severgnini} P.,   {Vignali} C.,  2024,
  \mn@doi [\mnras] {10.1093/mnrasl/slae033}, \href
  {https://ui.adsabs.harvard.edu/abs/2024MNRAS.531L..76Z} {531, L76}

\makeatother
\end{thebibliography}

%\begin{thebibliography}{99}
%\end{thebibliography}

% Don't change these lines
\bsp	% typesetting comment
\label{lastpage}
\end{document}